\documentclass[aps,prl,reprint,superscriptaddress]{revtex4-1}
\usepackage{epsfig}             
\usepackage{epstopdf}           
\usepackage{hyperref}           
\usepackage{color}
\usepackage{colortbl}
\usepackage{graphicx}
\usepackage{amssymb,amsmath}
\usepackage{subfig}
\usepackage{caption}
\usepackage{enumerate}
\usepackage{gensymb}
\usepackage{url}
\usepackage{floatrow}
\usepackage{gensymb}
\usepackage{multirow}

\begin{document}

\title[States of $^{12}$N with enhanced radii]{States of $^{12}$N with enhanced radii}

\author{A.S. Demyanova}
\affiliation{National Research Centre Kurchatov Institute, Moscow 123182, Russia}
\author{A.N. Danilov}
\affiliation{National Research Centre Kurchatov Institute, Moscow 123182, Russia}
\author{A.A. Ogloblin}
\affiliation{National Research Centre Kurchatov Institute, Moscow 123182, Russia}
\author{V.I. Starastsin}
\affiliation{National Research Centre Kurchatov Institute, Moscow 123182, Russia}
\author{S.V. Dmitriev}
\affiliation{National Research Centre Kurchatov Institute, Moscow 123182, Russia}
\author{W.H. Trzaska}
\affiliation{Department of Physics, University of Jyväskylä, P.O. Box 35, FI-40014 Jyväskylä, Finland}
\author{S.A. Goncharov}
\affiliation{Lomonosov Moscow State University, GSP-1, Leninskie Gory, Moscow 119991, Russia}
\author{T.L. Belyaeva}
\affiliation{Universidad Aut\'{o}noma del Estado de M\'{e}xico, C.P. 50000, Toluca, M\'{e}xico}
\author{V.A. Maslov}
\affiliation{Flerov Laboratory for Nuclear Research, JINR, Dubna, Moscow Region 141980, Russia}
\author{Yu.G. Sobolev}
\affiliation{Flerov Laboratory for Nuclear Research, JINR, Dubna, Moscow Region 141980, Russia}
\author{Yu.E. Penionzhkevich}
\affiliation{Flerov Laboratory for Nuclear Research, JINR, Dubna, Moscow Region 141980, Russia}
\author{S.V. Khlebnikov}
\affiliation{V.G. Khlopin Radium Institute, St. Petersburg 194021, Russia}
\author{G.P. Tyurin}
\affiliation{V.G. Khlopin Radium Institute, St. Petersburg 194021, Russia}
\author{N. Burtebaev}
\affiliation{Institute of Nuclear Physics, National Nuclear Center of Republic of Kazakhstan, Almaty 050032, Republic of Kazakhstan}
\author{D. Janseitov}
\affiliation{Flerov Laboratory for Nuclear Research, JINR, Dubna, Moscow Region 141980, Russia}
\affiliation{Institute of Nuclear Physics, National Nuclear Center of Republic of Kazakhstan, Almaty 050032, Republic of Kazakhstan}
\author{Yu.B. Gurov}
\affiliation{National Research Nuclear University MEPhI (Moscow Engineering Physics Institute), Moscow 115409, Russia}
\author{J. Louko}
\affiliation{Department of Physics, University of Jyväskylä, P.O. Box 35, FI-40014 Jyväskylä, Finland}
\author{V.M. Sergeev}
\affiliation{National Research Centre Kurchatov Institute, Moscow 123182, Russia}
\affiliation{Lomonosov Moscow State University, GSP-1, Leninskie Gory, Moscow 119991, Russia}


\date{\today}

\begin{abstract}
The differential cross sections of the $^{12}$C($^3$He,t)$^{12}$N reaction leading to formation of the 1$^+$ (ground state), 2$^+$(0.96 MeV), 2$^{-}$(1.19 MeV), and 1$^{-}$(1.80 MeV) states of $^{12}$N are measured at $E$($^3$He)=40 MeV. The analysis of the data is carried out within the modified diffraction model (MDM) and distorted wave Born approximation (DWBA). Enhanced $rms$ radii were obtained for the ground, 2$^{-}$(1.19 MeV), and 1$^{-}$(1.80 MeV) states. We revealed that $^{12}$B, $^{12}$N, and $^{12}$C in the IAS with T=1, and spin-parities 2$^{-}$ and 1$^{-}$ have increased radii and exhibit properties of neutron and proton halo states.

\end{abstract}

\maketitle

\section{Introduction}\label{intro}

Recently the evidence of the excited states of light nuclei with enlarged radii, located close to and above the particle emission threshold, was convincingly demonstrated (see, e.g., Ref.~\cite{Ogloblin2017} and references therein). The existence of neutron halos in the short-lived excited states of some stable and radioactive nuclei was revealed, in particular, by the asymptotic normalization coefficient (ANC) analysis of the neutron-transfer reactions ~\cite{Liu2001, Belyaeva2014}. Thus Liu et al. ~\cite{Liu2001} analyzing the data on the (d,p) reactions on $^{12}$C and $^{11}$B by the ANC method reported the observation of halos in the first 1/2$^+$(3.09 MeV) excited state of $^{13}$C, and the second 2$^-$(1.67 MeV) and third 1$^-$(2.62 MeV) excited states of $^{12}$B. 

Similar results were obtained by the ANC analysis of the $^{11}$B(d,p)$^{12}$B reaction at $E_{lab}$ = 21.5 MeV carried out in our group ~\cite{Belyaeva2018}. Radii of the valence neutron for the first five excited states of $^{12}$B were determined. Calculations showed that the $rms$ radii of the last neutron in the second 2$^-$(1.67 MeV) and the third 1$^-$(2.62 MeV) excited states of $^{12}$B far exceed those for the ground state (g.s.) and the first 2$^+$(0.95 MeV) excited state. Exactly, for the 2$^-$ state, the excess is a factor of 1.7, and for the 1$^-$ state, it is a factor of 2.1, with respect to the $rms$ radius of the ground state. Moreover, a probability of the last neutron to be outside the range of the interaction radius, so-called D$_1$ coefficient, was obtained to be 53$\%$ and 62$\%$, respectively. It should be noted that a formal criterion of a halo state is that D$_1$ should be more than 50$\%$ and it is fulfilled in both cases. 

Accordingly to charge independence of nuclear forces, mirror nuclei are isobars that have proton and neutron numbers interchanged. Some states of mirror nuclei with the same quantum numbers (isospin, spin/parity), isobaric analogue states (IAS), can form the isospin or isotopic multiplets (doublets, triplets, etc.) and then approximately have the same structure and radii.

Natural question arises: what we can expect in the IAS of $^{12}$B in the mirror $^{12}$N nucleus? The IAS with probable exotic structure are marked by thick lines in Fig. 1. The IAS that presumable have halos are determined in a more complicated manner: replacing the neutron in the halo state with a proton does not necessarily lead to the appearance of a similar proton structure. The fact is that the appearance of a halo is determined by the proximity of the valence nucleon to the emission threshold, and it can be very different for a neutron and a proton. One notable example is the IAS of mirror $^{13}$C and $^{13}$N nuclei. $^{13}$C in the 1/2$^+$, 3.09-MeV state has a neutron halo ~\cite{Liu2001,Belyaeva2014} that satisfies all halo criterions. The 1/2$^+$, 2.37-MeV IAS in $^{13}$N does not lie in the discrete spectrum, but in the continuum spectrum, and therefore the proton wave function differs from the neutron one. An increase of the $^{13}$N radius in this state is also observed ~\cite{Demyanova2016}, but halo criterions for the continuum states for the present are not clearly formulated.

\begin{figure*}
	\centering	
	\includegraphics[width=\textwidth,  height=6cm]{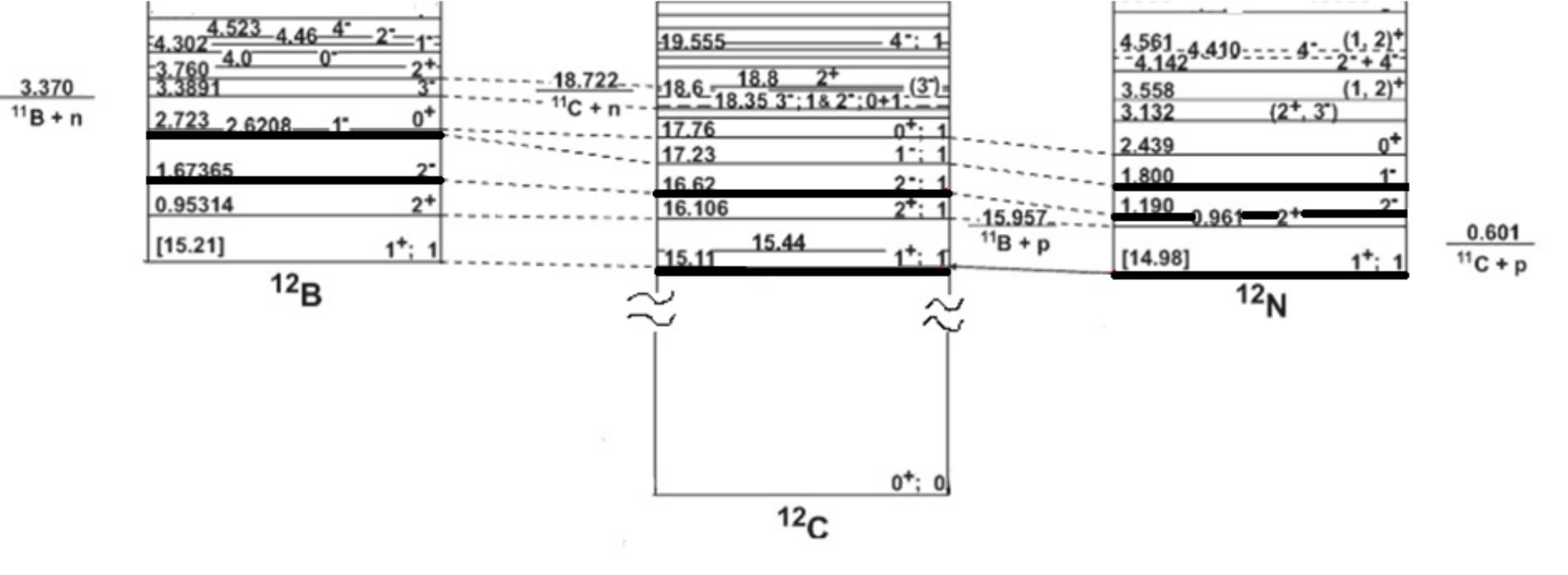}
	\caption{Triplet A=12. The thick lines indicate possible states with increased radii.}
	\label{Intro-fig:sketch}
\end{figure*}

Some evidences of a proton halo were already found in the ground state of $^{12}$N ~\cite{Li2010,Warner1998}, where the last proton has a binding energy $S$ of only 0.6 MeV. This state satisfies a criterion (necessary, but not sufficient) given by P. G. Hansen et al. ~\cite{Hansen1995} for halo existence: $SA^{2/3} \approx$ 2 to 4 MeV.

Now we study excited states of $^{12}$N, namely the 2$^+$(0.96 MeV), 2$^-$(1.19 MeV), and 1$^-$(1.80 MeV) states of $^{12}$N. We propose to use the MDM and apply it to analyze the ($^3$He,t) reaction data. Obtained radii for $^{12}$N in the 2$^-$ and 1$^-$ states will be compared with those received for the excited states of $^{12}$B ~\cite{Belyaeva2018}. The problem is that existing data are not completed enough to make definite conclusion about the radii of the 2$^-$ and 1$^-$ states in $^{12}$N. The existing in the literature data are presented only at three energies: 36 ~\cite{Artemov1970}, 49.8 ~\cite{Ball1969}, and 81 MeV ~\cite{Sterrenburg1983}. The data at 36 MeV contain only the angular distributions for the g.s. and the 0.96-MeV states. The data at 49.8 MeV contain the angular distributions for the g.s., 0.96-MeV, and 1.20-MeV states. The data at 81 MeV contain all interested for us states, but they present only one indinstinct oscillation in the angular distributions. The angular distribution for the 0.96-MeV state obtained at 81 MeV ~\cite{Sterrenburg1983} is not comparable with others, if it would be drawn as a function of linear transferred momentum. This fact stimulate us to carried out a new experiment on the $^{12}$C($^3$He,t)$^{12}$N reaction at $E$($^3$He) = 40 MeV.

\section{EXPERIMENTAL PROCEDURE AND RESULTS}\label{intro}

The measurements were conducted at the University of Jyväskylä (Finland) using the K130 cyclotron ~\cite{Trzaska2018} to produce a $^3$He beam at $E$($^3$He) = 40 MeV. The 150 cm diameter Large Scattering Chamber was equipped with four $\Delta E-E$ detector telescopes, each containing two independent $\Delta E$ detectors and one common $E$ detector. So each device allowed carrying out measurements at two angles. The measurements in c.m. angular range 10$\degree$ were conducted in one exposure. The differential cross sections of the $^{12}$C($^3$He,t)$^{12}$N  reaction were measured in the c.m. angular range of 8$\degree$-69$\degree$. Self-supported $^{12}$C foils of 0.23 and 0.5 mg/cm$^2$ thicknesses were used as targets. The beam intensity was about 20 particle nA.

It should be mentioned that, before starting measurements, beam monochromatization was done ~\cite{Trzaska2018}, which made it possible to diminish beam energy spreading up to three times and obtain a total energy resolution about 140 keV. In Fig. 2, a sample spectrum from the $^{12}$C($^3$He,t)$^{12}$N reaction at $\theta_{lab}$ = 28\degree  showing the excitation of the $^{12}$N states up to $E_{x} \approx$ 2.5 MeV is presented. All peaks are distinct and are well separated, except the 0.96 and 1.19-MeV states. The beam monochromatization allows us to separate these neighboring 0.96-MeV 2$^{+}$ and 1.19$^{-}$ MeV 2$^{-}$ states of $^{12}$N.

\begin{figure}
\centering	
\includegraphics[width=.85\columnwidth]{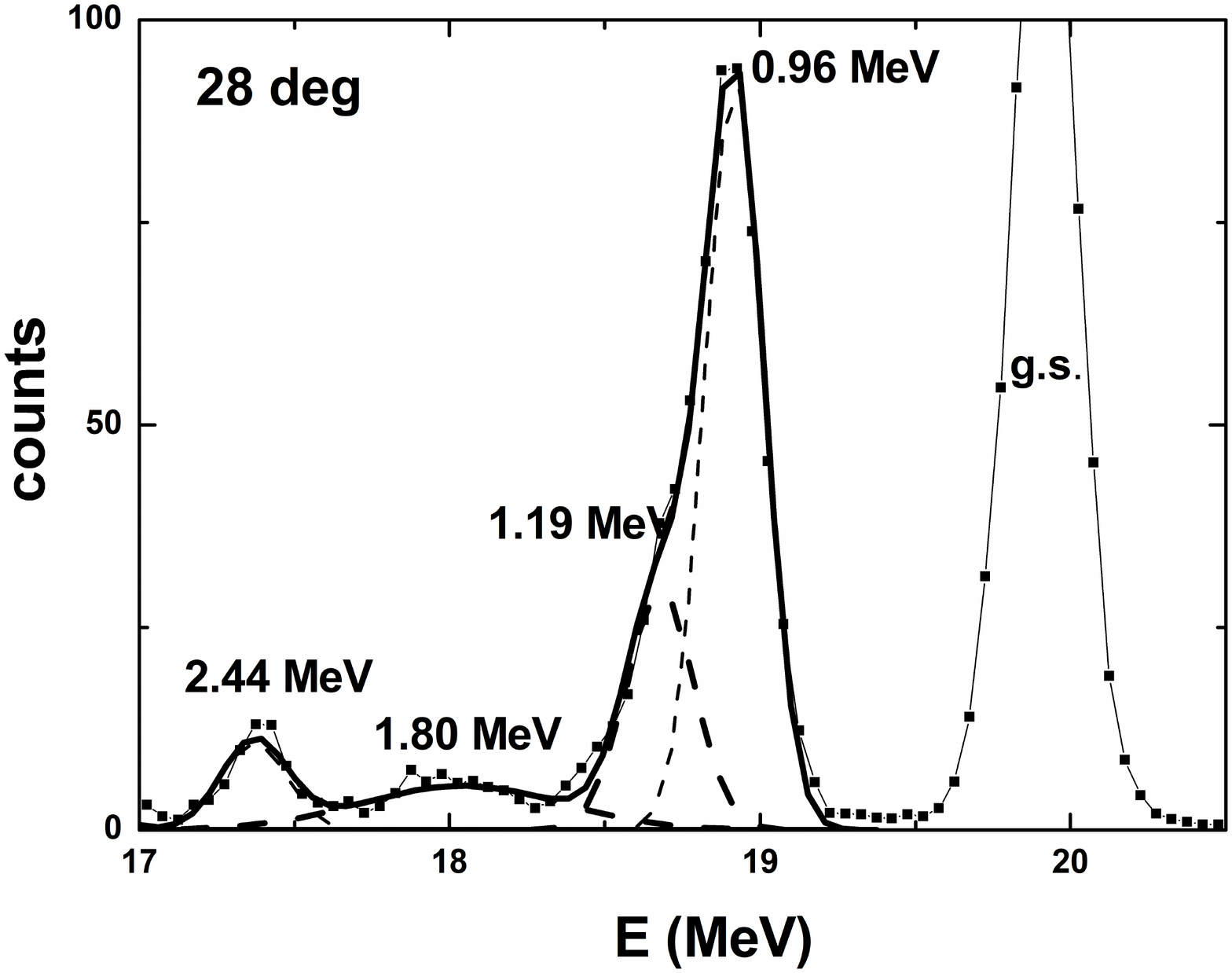}
\caption{Tritium spectrum from the $^{12}$C($^{3}$He,t)$^{12}$N  reaction at $\theta_{lab}$ = 28$\degree$ with the excitation of the $^{12}$N states up to $E_{x}$ $\approx$ 2.5 MeV.}
\label{Intro-fig:sketch}
\end{figure}

A standard expansion method was employed to obtain cross sections: the spectrum peaks were fitted with a Gaussian shape. The peak positions and widths were fixed in accordance with the world-average values, and the area under the peak was the only free parameter.

Triton angular distributions for the g.s. and three first excited states of $^{12}$N: 0.96-MeV 2$^{+}$, 1.19-MeV 2$^{-}$, and 1.80-MeV 1$^{-}$ were measured. The resulting differential cross sections for the $^{12}$C($^{3}$He,t)$^{12}$N reaction with DWBA calculations are presented in Fig. 3.

\section{DWBA ANALYSIS}\label{intro}

The cross sections for the studied charge exchange reaction were calculated by the DWBA using the DWUCK4 code  ~\cite{Kunz1993}.

For the entrance and exit channels, the semi-microscopic optical potential was used in the framework of the dispersion optical model SMDOM, where the mean field potentials were calculated in the double folding model (for details of the formalism, see ~\cite{Kunz2007,Goncharov2018}). For the target $^{12}$C nucleus and the final $^{12}$N nucleus, we used the empirical density of the Fermi form, which gives the correct $rms$ matter radius. Density distributions for nuclei with a mass number A = 3 were calculated through the charge form factors of these nuclei, as in Ref. ~\cite{Goncharov1996}. From the analysis of the available data (from the EXFOR BNL database http://www.nndc.bnl.gov/exfor/exfor.htm) of the $^3$He + $^{12}$C elastic scattering in the energy range from 24 to 217 MeV, the parameters of the dynamic polarization potential (DPP) were found. Reasonable reaction cross sections and a good description of the angular distributions of the $^3$He + $^{12}$C elastic scattering in the energy range from 50 to 217 MeV were obtained. For lower energies, only the angle region up to 70$\degree$ was well described. The description of angular distributions is better than that given by the known global potentials ~\cite{Liang2009}.

For the exit channel, t + $^{12}$N, we used the DPP parameters as for the $^3$He + $^{12}$C channel estimated at the corresponding energies.

The microscopic approach (implemented in DWUCK4 for calculating inelastic processes) was used to describe the charge exchange reaction. Here, the radial form factor was determined by the overlap integral of the single-particle wave function (with quantum numbers $nlj$) of the neutron in the target $^{12}$C and the single-particle wave function (with quantum numbers $n'l'j'$) of the proton in the final nucleus $^{12}$N with interaction including central and tensor components, for which the Yukawa potential shape was chosen. We considered single-particle configurations $(nlj,n'l'j')$, which give acceptable values of the transferred moments $LSJ$. The single-particle wave functions of the target and the bound states of the final nucleus normalized to unity were calculated using the standard procedure for adjusting the well depth for fixed nucleon separation energy. The geometric parameters, radius $r_{0}$ and diffuseness $a$, of the Woods-Saxon single-particle potential were considered as adjusted parameters. Note that all excited states of $^{12}$N belong to the continuous spectrum. For these states, the procedure proposed in ~\cite{Vincent1973} and implemented in the DWUCK4 code was applied, assuming that a depth $V$ of the one-particle potential is also adjusted parameter.

Calculations of the differential cross section include the coherent contribution of the amplitudes corresponding to various combinations of transferred moments. The parameters of the inverse radii of the interaction ($\mu_{\tau}$,  $\mu_{\sigma\tau}$, $\mu_{T\tau}$), their strengths ($\nu_{\tau}$, $\nu_{\sigma\tau}$, $\nu_{T\tau}$), and the normalizations of the contributions of the given configurations $N_{LSJ}(nlj,n'l'j')$ were also the free parameters. The latter, in fact, are the product of the squares of single-particle spectroscopic amplitudes (or single-particle widths) and other factors that are not calculated in the DWUCK4 code. The parameters were selected to describe the experimental differential cross sections, at least in the region of the forward angles (in the region of the “main maximum”).

In all cases, the fixed values of $\mu_{\tau}$ = 0.9 and $\mu_{\sigma\tau}$ = $\mu_{T\tau}$, = 0.7 fm$^{-1}$ were used, as well as $\nu_{\tau}$ = +7, $\nu_{\sigma\tau}$ = -3,  and  $\nu_{T\tau}$ = -9  MeV. In the g.s. of $^{12}$C for a single-particle neutron state $nlj = 1p_{3/2}$ the geometrical parameters $r_0$ = 1.25 fm and $a$ = 0.65 fm were taken. Table 1 shows the values of all other parameters adjusted to describe the experimental differential cross sections in the region of the “main maximum”.

Note that the single-particle $1p_{1/2}$ proton wave function for the g.s. of $^{12}$N has the $rms$ radius of 7.2 fm. If we estimate the radius of the $^{11}$C core equal to 2.2 fm, this value leads using relationship proposed by Tostevin and Al-Khalili ~\cite{Tostevin1997} to the $rms$ radius of $^{12}$N in the 1$^+$ g.s. equal to about 2.9 fm.

\begin{table}
\caption{Parameters $V$, $r_0$ and $a$ used in the form factor calculations and relative norms $N_{LSJ}$.}
\begin{tabular}{|l|l|l|l|l|l|l|}
 \hline
 States $^{12}$N & $V$, MeV & $r_0$, fm & $a$, fm  & $n’ l’j’$ &  $L S J$  & $N_{LSJ}$\\
 \hline
 \multirow{2}{*}{g.s (1$^+$)} & \multirow{2}{*}{24.7} & \multirow{2}{*}{1.25} & \multirow{2}{*}{2.50} & \multirow{2}{*}{1$p_{1/2}$} & 0     1     1 & 3.24\\
 & & & & & 2     1     1 & 0.81\\
 \hline
0.96 (2$^+$) & 86.0 & 1.40 & 0.35 &  1$p_{3/2}$ & 2     0     2    & 3.1\\
\hline
\multirow{2}{*}{1.19 (2$^-$) } & \multirow{2}{*}{46.0} & \multirow{2}{*}{1.20} & \multirow{2}{*}{1.00} & \multirow{2}{*}{1$d_{5/2}$} & 1      1    2 & 1.90\\
 & &  &  &   & 3      1    2 & 4.75\\
 \hline
\multirow{2}{*}{1.8 (1$^-$)} & \multirow{2}{*}{59.0} & \multirow{2}{*}{1.40} & \multirow{2}{*}{0.20} & \multirow{2}{*}{1$d_{3/2}$}  & 1     0     1 & 8.1\\
 &  & &  &   & 1     1     1 & 8.1\\
 \hline
\end{tabular}
\end{table}

Our analysis showed that with a different choice of the parameters of the Yukawa interaction, the single-particle Woods-Saxon potential and normalizations of the contributions of the given configurations, one can also obtain a somewhat worse, but satisfactory description in the region of the main maximum of the differential cross section, while other values are obtained for estimating the $rms$ radius of the nucleus $^{12}$N in the 1$^+$ g.s. So, the uncertainty of the estimation can reach about 15-20\%. The presented set of parameters, in our opinion, gives a better description, including the data mentioned above at other energies.

\begin{figure}
	\centering	
	\includegraphics[width=.85\columnwidth]{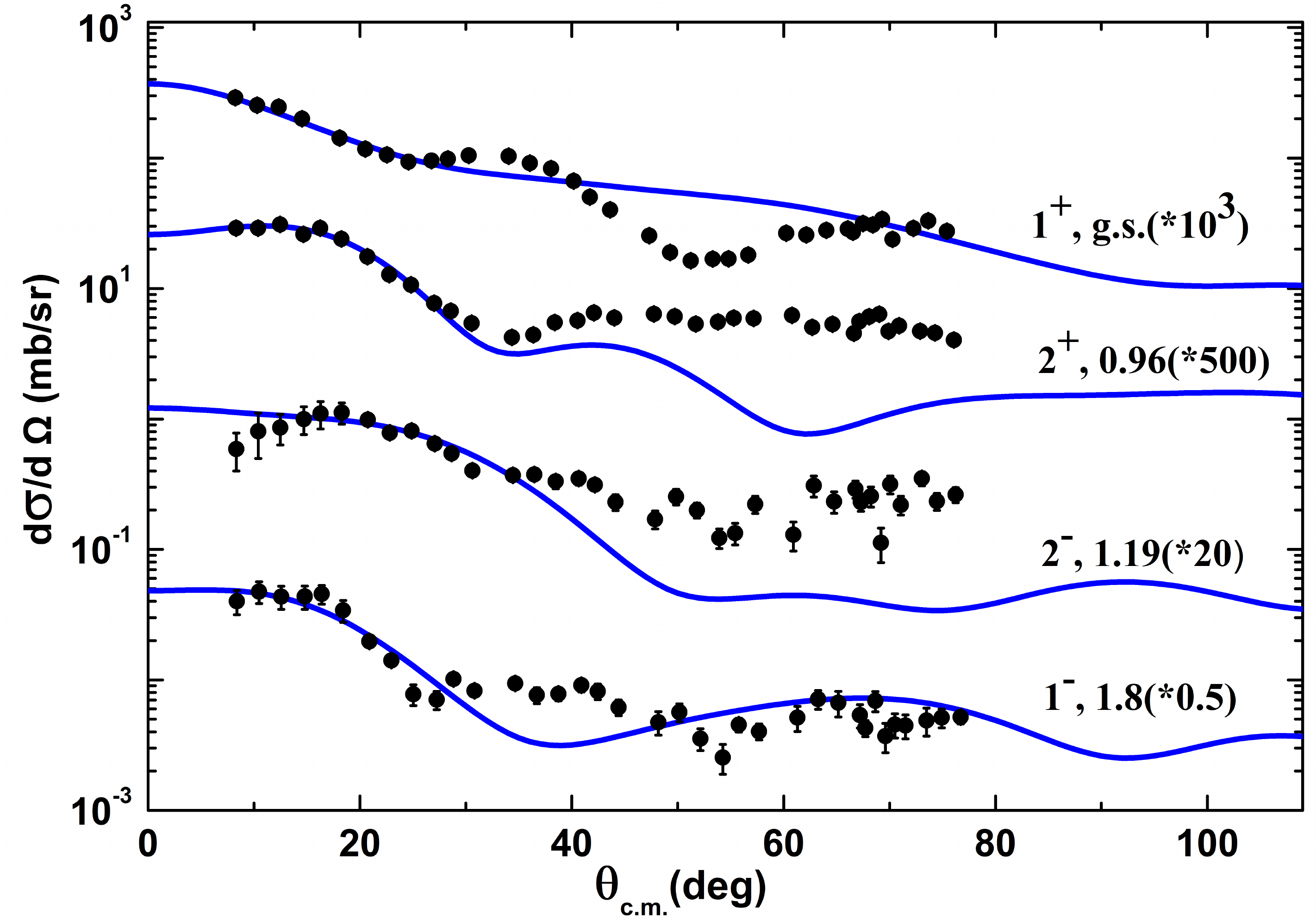}
	\caption{Triton angular distributions from the $^{12}$C($^{3}$He,t)$^{12}$N reaction at E($^{3}$He) = 40 MeV populated the 1$^{+}$( g.s), 2$^{+}$(0.96 MeV), 2$^{-}$(1.19 MeV), and 1$^{-}$(1.80 MeV) states of $^{12}$N. The curves correspond to the DWBA calculations.}
	\label{Intro-fig:sketch}
\end{figure}

\section{MDM ANALYSIS}\label{intro}

In our group, several methods are developed to be used for measuring radii of nuclei in the short-lived excited states: the MDM ~\cite{Danilov2009}, the ANC method  ~\cite{Liu2001,Belyaeva2014}, and the nuclear rainbow method (NRM) ~\cite{Demyanova2008}. The ANC method is the most appropriate to measure halo radii. Unfortunately, the ANC method is applicable only to bound states. In order to measure nuclear radii in unbound states, we propose to use the MDM for the analysis of inelastic differential cross sections. However, not all excited states can be populated through inelastic scattering. It has been known for a long time that charge exchange reactions have much in common with inelastic scattering ~\cite{Lane1962}. We therefore propose to extend the MDM for measuring radii of nuclei in the proton rich states (for performing proton-halo searches) by analysis of the ($^{3}$He,t)  reaction data. 

Let us briefly consider the main aspects of this approach ~\cite{Demyanova2017}. In the plane wave approximation, the cross section of a charge exchange reaction is described by the spherical Bessel functions (in the case of scattering, by cylindrical ones), so: 
\begin{equation}
\frac{d\sigma}{d\Omega} \sim [j_L(qR)]^2
\end{equation}
where $q$ is the linear transferred momentum and $R$ is a radial parameter named diffraction radius. In accordance with (1), the first small-angle minima (maxima) of experimental angular distributions are associated with the squared extrema of the Bessel function of corresponding order depending on the linear transferred momentum. So, diffraction radius as the only parameter of the model can be determined.

Direct application of the MDM would involve a comparison of the inelastic and elastic scattering: 
\begin{equation}
\begin{gathered}
<R^*(^{12}N)>=<R_0(^{12}N)>+ \\
[R^*_{dif}(^{12}N)-R_{dif}(^{12}N)]
\end{gathered}
\end{equation}

 where $<R_0>$ is a presumably known $rms$ radius of the target nucleus in the g.s., $R^*_{dif}$ and $R_{dif}$ are the diffraction radii determined from the positions of the minima and maxima of the experimental angular distributions of the scattering leading to the excited state and the g.s., respectively.
 
 By analogy with the scattering, in charge exchange reactions, the $rms$ radius of the nucleus in the excited state is estimated using practically the same formula. As for the ground-state diffraction radius, $R_{dif}$, it should be determine in accordance with the MDM procedure, in the case of the $^{12}$C($^3$He,t)$^{12}$N reaction, from elastic scattering of $^3$He or $^3$H on $^{12}$N, but, as mentioned above, this is almost impossible. Therefore, it is proposed to use elastic $^3$He + $^{12}$C scattering. Since the radii of the ground states have close values, $<R_0(^{12}C)>$ = 2.35 $\pm$ 0.02 fm  ~\cite{Ozawa2001}  versus  $<R_0(^{12}N)>$ = 2.47 $\pm$ 0.07 fm ~\cite{Ozawa2001}, their diffraction radii should only differ by a correction, which takes into account the fact that the Coulomb interaction in the exit channels is different for the triton and $^{3}$He nucleus: 
 
\begin{equation}
\begin{gathered}
<R^*(^{12}N)>=<R_0(^{12}N)>+ \\ [R^*_{dif}(^{12}N)-R'_{dif}(^{12}N)]
\end{gathered}
\end{equation}
According to Ref. ~\cite{Satchler1990}, the “corrected” ground-state diffraction radius for $^{12}$N is:
 \begin{equation}
 \begin{gathered}
<R'(^{12}N)>=\frac{\eta}{k} + \{[R_{dif}(^{12}C)]^2+[\frac{\eta}{k}]^2\}^{\frac{1}{2}}, \\ \frac{\eta}{k} = \frac{Z_1Z_2e^2}{2E}
\end{gathered}
\end{equation}
There are convincing arguments to apply the MDM to charge exchange reactions in order to study the IAS of mirror nuclei. We have first applied this approach to determine the proton halo in the unbound state of $^{13}$N ~\cite{Demyanova2016}. More detailed description is present in ~\cite{Demyanova2017}. 

\section{RESULTS AND DISCUSSION}\label{intro}

Let us discuss the results of the MDM analysis to the existing and our new data on the $^{12}$C($^3$He,t)$^{12}$N reaction at 40 MeV. 
Angular distributions of the $^{12}$C($^3$He,t)$^{12}$N reaction with excitation of the 1$^+$ g.s. and the excited 1.19-MeV 2$^-$, and 1.80-MeV 1$^-$ states of $^{12}$N at incident energies 40, 49.8, and 81 MeV are presented in Fig. 4. The cross sections of the inelastic $^{3}$He + $^{12}$C scattering at E($^{3}$He) = 49.8 MeV with excitation of the IAS in $^{12}$C are also shown. Arrows correspond to the positions of extrema used for the MDM analysis in accordance with (1). 

Comparison of the data indicate that the positions of minima and maxima of angular distributions of the ($^3$He,t) reaction at different energies (marked by arrows) move to smaller angles with an energy increase. Moreover, the angular distributions of the ($^{3}$He,t) reaction at 49.8 MeV are coincident with those for the inelastic $^{3}$He scattering on $^{12}$C. This confirms a diffraction nature of these extrema.

\begin{figure}
	\centering	
	\includegraphics[width=.85\columnwidth]{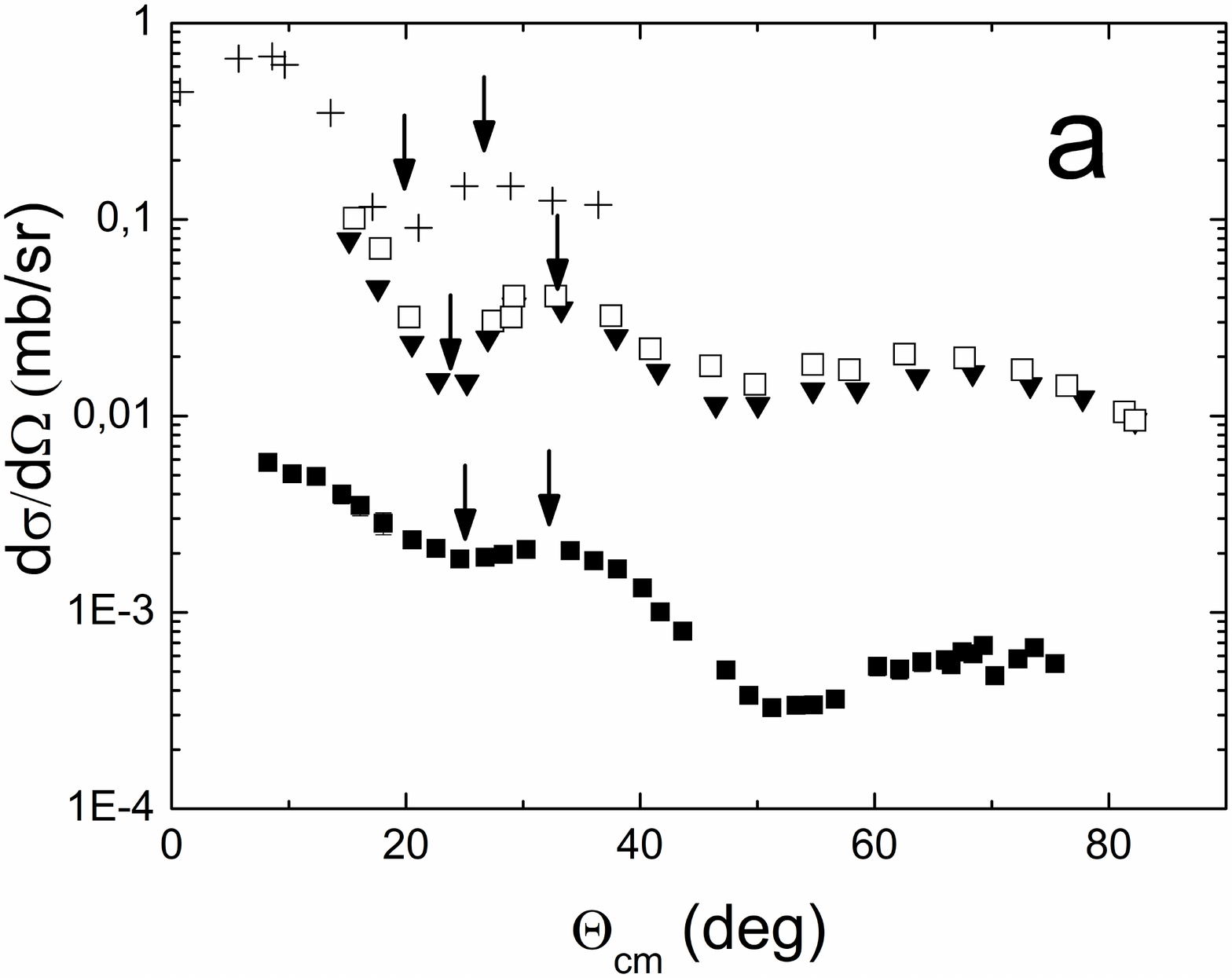}
        \includegraphics[width=.85\columnwidth]{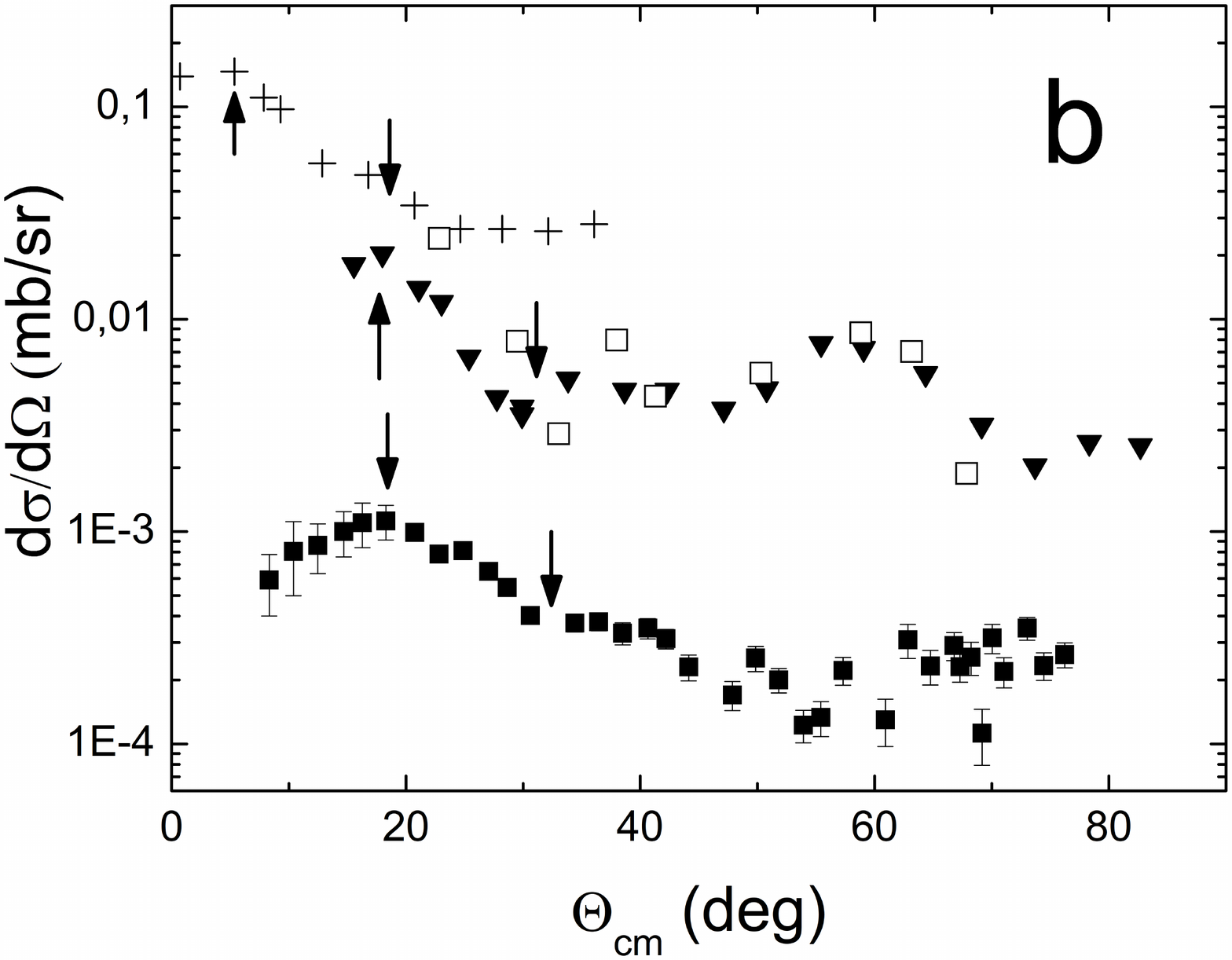}
       \includegraphics[width=.85\columnwidth]{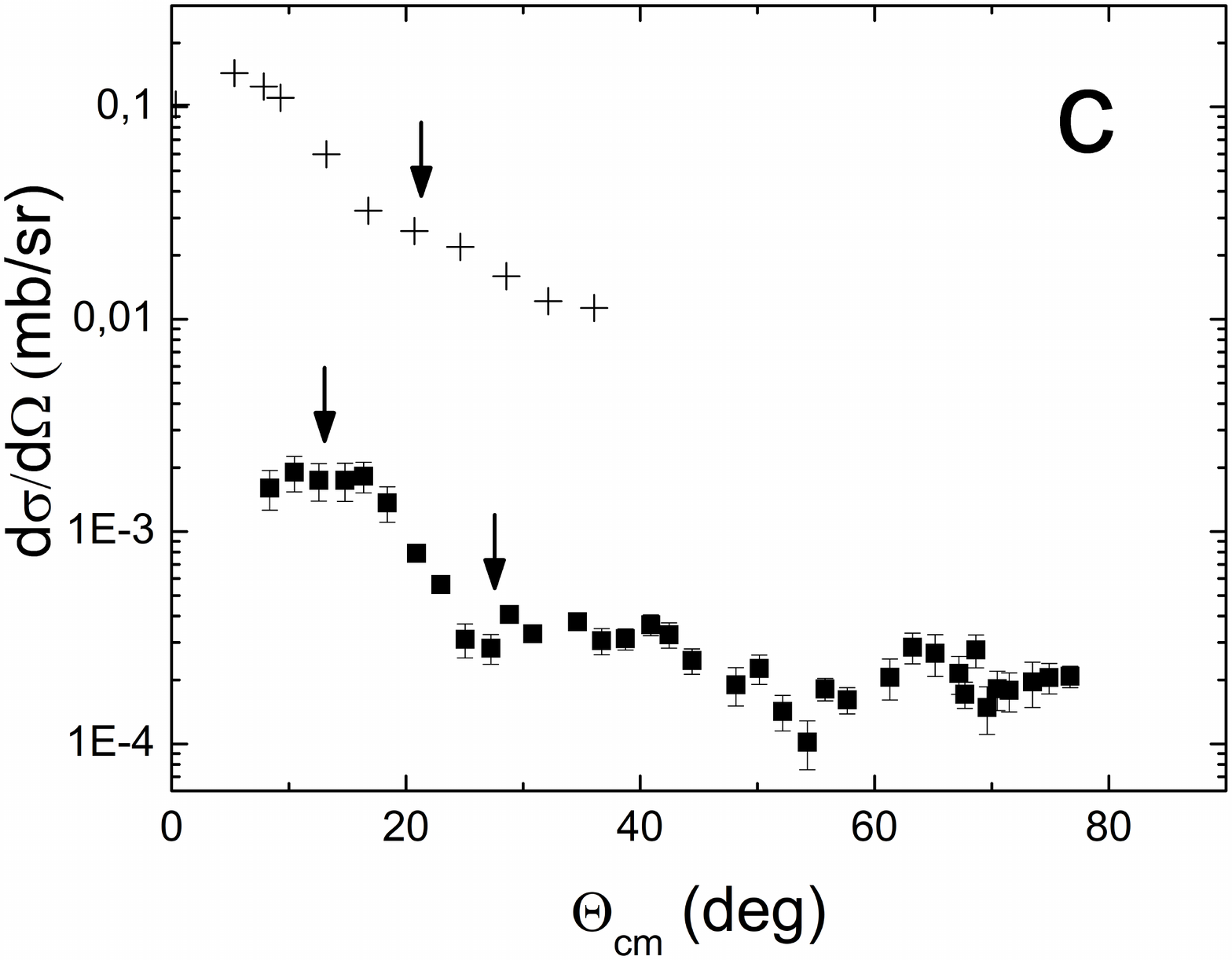}
	\caption{Triton angular distributions from the $^{12}$C($^3$He,t)$^{12}$N reaction with the excitation of (a) the 1$^{+}$ g.s., (b) the 1.19-MeV 2$^{-}$ state, and (c) the 1.80-MeV 1$^{-}$  state of $^{12}$N at 40 MeV (black squares, the cross sections are multiplied by a factor of 1/50), 49.8 MeV (black triangles, the cross sections are multiplied by a factor of 1/5), and 81 MeV (crosses). Opened squares correspond to the cross sections of inelastic scattering of $^{3}$He on $^{12}$C with the excitation of corresponding isobaric analogue states in $^{12}$C (multiplied by a factor of 19/50).}
	\label{Intro-fig:sketch}
\end{figure}

In accordance with Eq. (4), we take, as the starting point, the diffraction radius of the elastic $^{3}$He + $^{12}$C scattering. Due to the fact that the $^{12}$C($^{3}$He,t)$^{12}$N reaction has $Q$ value of -17.36 MeV, we use the elastic scattering in energy interval of 20-25 MeV and determine $R_{dif}(0)$ = 5.45 $\pm$ 0.17 fm based on the data at 25.3 MeV ~\cite{Matsuda1968}. Then the “corrected” ground-state diffraction radius for $^{12}$N becomes: $R'_{dif}$ = 5.8 $\pm$ 0.2 fm. We try to make estimations of $rms$ radius of the g.s of $^{12}$N. In this case, it can’t be taken as $R_0$ in (3) and should be determined. We propose to use radius of $^{12}$C 2.35 fm ~\cite{Ozawa2001}. Then an estimate of the $rms$ radius of $^{12}$N in the g.s. gives 2.8 $\pm$ 0.4 fm, which is consistent with the estimates resulting from the DWBA analysis.

The diffraction radii for the 1.19 MeV and 1.80 MeV states are determined by using (1). The 1.19 MeV state is excited by transfer of two angular momentums $L$=0 and $L$=2 and corresponding diffraction radii are found to be 6.0 $\pm$ 0.1 and 6.3 $\pm$ 0.3 fm, respectively. The diffraction radius for the 1.80 MeV state is 6.3 $\pm$ 0.1 fm. These values are larger than $R'_{dif}$, so the $rms$ radii in these states are increased. Obtained preliminary $rms$ radii are present in Table 2. 

The diffraction and $rms$ radii of $^{12}$C in the IAS were determined by the MDM from the inelastic $^{3}$He + $^{12}$C scattering. Within the error bars, the $rms$ radii of $^{12}$C in the 15.11-MeV 1$^{+}$ and the 16.57-MeV 2$^{-}$ states agree with the $rms$ radii of the IAS of $^{12}$N. The ANC analysis gives approximately the same radii ~\cite{Belyaeva2020}. Moreover, D$_1$ coefficient for the 2$^{-}$ state is more than 50\%, which indicate that the 16.57-MeV 2$^-$  state of $^{12}$C can be considered as a proton halo-like state. Complete results of the ANC analysis will be published later ~\cite{Belyaeva2020}. 

Summary of preliminary results of the MDM analysis in comparison with the results for $^{12}$B ~\cite{Belyaeva2018} and preliminary results of ANC analysis for $^{12}$C ~\cite{Belyaeva2020} are present in Table 2.

\begin{table}[h]
\caption{Preliminary $rms$ radii of $^{12}$N obtained by the MDM analysis of the $^{12}$C($^3$He,t)$^{12}$N reaction in comparison with the $rms$ radii of $^{12}$B ~\cite{Belyaeva2018} and $^{12}$C ~\cite{Belyaeva2020}.}
\begin{tabular}{|c||c|c|c||c|c|}
\hline
J$^{\pi}$	& $^{12}$B state& 	R$_{rms}$ & D$_1$ &  $^{12}$N state & R$_{rms}$ \\
& (MeV) & (fm) & (\%) & (MeV) & (fm) \\
\hline
& & & & & 2.47 $\pm$ 0.07 $^{a)}$ \\
1$^{+}$ & g.s & 2.39$\pm$0.02$^{a)}$ & 11$^{b)}$ & g.s & 2.5 $^{c)}$ \\ 
& & & & & 2.8 $\pm$ 0.4 \\
\hline
2$^{-}$ & 1.67 & 2.58 $\pm$ 0.11 $^{b)}$ & 53 $^{b)}$ & 1.19 & 2.8 $\pm$ 0.3 \\
\hline
1$^{-}$ & 2.62 & 2.86 $\pm$ 0.11 $^{b)}$ & 62 $^{b)}$ & 1.8 & 3.0 $\pm$ 0.1 \\
\hline
J$^{\pi}$ & $^{12}$C state & R$_{rms}$ & D$_1$ & \multicolumn{2}{c}{} \\
& (MeV) & (fm) & (\%) & \multicolumn{2}{c}{}\\
\cline{1-4}
1$^{+}$ & 15.11 & 2.60 $\pm$ 0.06$^{d)}$ & 35$^{d)}$ & \multicolumn{2}{c}{}\\
\cline{1-4}
2$^{-}$ & 16.57 & 2.85 $\pm$ 0.06$^{d)}$ & 50$^{d)}$ & \multicolumn{2}{c}{}\\
\cline{1-4}

\multicolumn{6}{l}{Notes.$^{a)}$~\cite{Ozawa2001}; $^{b)}$~\cite{Belyaeva2018}; $^{c)}$~\cite{Li2010}; $^{d)}$~\cite{Belyaeva2020}.}
\end{tabular}
\end{table}

\section{Conclusions}\label{intro}
The differential cross sections of the $^{12}$C($^3$He,t)$^{12}$N reaction leading to formation of the 1$^+$ (ground state), 2$^+$ (0.96 MeV), 2$^{-}$ (1.19 MeV), and 1$^{-}$ (1.80 MeV) states of $^{12}$N are measured at E($^{3}$He) = 40 MeV. The analysis of the data is carried out within the Modified Diffraction Model and Distorted Wave Born Approximation. Enhanced $rms$ radii were obtained for the 2$^{-}$ (1.19 MeV) and the 1$^{-}$ (1.80 MeV) states of $^{12}$N. The MDM analysis and DWBA analysis have showed that the $rms$ radius of $^{12}$N in the ground state is also enlarged. Preliminary ANC analysis showed that the isobaric analogue states in $^{12}$C: 1$^{+}$ at 15.11 MeV and 2$^{-}$ at 16.57 MeV, also have increased radii and a large value of D$_1$ coefficients in the 2$^{-}$ states of $^{12}$C. Finally, we revealed that $^{12}$B, $^{12}$N, and $^{12}$C in the IAS with T = 1, and spin-parities 2$^{-}$ and 1$^{-}$ have increased radii and exhibit properties of neutron and proton halo states. 

\section{Acknowledgments}

The work was supported by the Russian grant RSF 18-12-00312.

\end{document}